\documentclass[final,5p,times,twocolumn]{elsarticle}
\usepackage{multicol}
\usepackage{graphicx}

\usepackage{array}
\usepackage{atlasphysics}

\usepackage{booktabs}
\usepackage{xspace}
\usepackage{rotating}
\usepackage{lscape}
\usepackage{longtable}
\usepackage{multirow}
\usepackage{float}
\usepackage{placeins}
\usepackage{mathrsfs}
\usepackage{amsmath,amssymb,latexsym}

\RequirePackage[colorlinks,breaklinks]{hyperref}
\hypersetup{linkcolor=blue,citecolor=blue,filecolor=black,urlcolor=blue}
\usepackage{hypernat}

\biboptions{numbers,square,comma,sort&compress}
\journal{Physics Letters B}

\newcommand{\lumi}{{20.3}}

\newcommand{\etmiss}{$E^{\textrm{miss}}_{\rm T}$}
\newcommand{\roofit}{\texttt{RooFit}}
\newcommand{\powheg}{\texttt{POWHEG-BOX}}
\newcommand{\pythia}{\texttt{PYTHIA}}
\newcommand{\alpgen}{\texttt{ALPGEN}}
\newcommand{\srhm}{SR$_\textrm{HM}$}
\newcommand{\srlm}{SR$_\textrm{LM}$}
\newcommand{\jimmy}{\texttt{JIMMY}}
\newcommand{\herwig}{\texttt{HERWIG}}
\newcommand{\sherpa}{\texttt{SHERPA}}
\newcommand{\mcatnlo}{\texttt{MC@NLO}}

\newcommand{\madgraph}{\texttt{MadGraph}}

\usepackage{preprintcover}  % load package

\PreprintCoverPaperTitle{Search for $WZ$ resonances in the fully leptonic channel using
  $pp$ collisions at $\sqrt{s} = 8$ \TeV~with the ATLAS detector}  % paper title

\PreprintIdNumber{CERN-PH-EP-2014-094}  % CERN preprint number

\PreprintCoverAbstract{A search for resonant $WZ$ production in the $\ell\nu\ell'\ell'$ ($\ell, \ell' = e,\mu$) decay channel
using 20.3 fb$^{-1}$ of $\sqrt{{s}} = 8$ \TeV~$pp$ collision data collected by the ATLAS
experiment at LHC is presented. No significant deviation from the Standard Model prediction is
observed and upper limits on the production cross sections of $WZ$ resonances from an extended gauge model
$W^{\prime}$ and from a simplified model of heavy vector triplets are
derived. A corresponding observed (expected) lower mass limit of 1.52
(1.49) \TeV~is derived for the $W^{\prime}$ at the 95\% confidence level.
}  % paper abstract
\PreprintJournalName{Physics Letters B}  % journal name
\begin{document}

\begin{frontmatter}

\title{Search for $WZ$ resonances in the fully leptonic channel using
  $pp$ collisions at $\sqrt{s} = 8$ \TeV~with the ATLAS detector}
%\author{G.~Aad \it{et al.} \\ (ATLAS Collaboration)}
\author{The ATLAS Collaboration}

  \begin{abstract}
    A search for resonant $WZ$ production in the $\ell\nu\ell'\ell'$ ($\ell, \ell' = e,\mu$) decay channel
using 20.3 fb$^{-1}$ of $\sqrt{{s}} = 8$ \TeV~$pp$ collision data collected by the ATLAS
experiment at LHC is presented. No significant deviation from the Standard Model prediction is
observed and upper limits on the production cross sections of $WZ$ resonances from an extended gauge model
$W^{\prime}$ and from a simplified model of heavy vector triplets are
derived. A corresponding observed (expected) lower mass limit of 1.52
(1.49) \TeV~is derived for the $W^{\prime}$ at the 95\% confidence level.

  \end{abstract}

\end{frontmatter}

\section{Introduction}
\label{introduction}

The search for diboson resonances is an essential complement to the
investigation of the source of electroweak symmetry breaking.
%Despite the apparent compatibility between the properties of the newly observed particle at the
Despite the compatibility between the properties of the newly discovered particle at the
LHC~\cite{ATLAS_HIGGS,CMS_HIGGS,ATLAS_SPIN,CMS_SPIN} with those expected for the Standard Model (SM)
Higgs boson, 
the naturalness problem associated with a light Higgs boson suggests that the SM is likely to be an effective theory valid only at low energies. 
Extensions of the SM, such as Grand Unified Theories~\cite{Langacker:1984dc}, 
Little Higgs models~\cite{ArkaniHamed:2002qy}, 
Technicolor~\cite{Lane:2002sm,Eichten:2007sx,Sannino:2004qp,Belyaev:2008yj}, 
more generic Composite Higgs models~\cite{Agashe:2004rs,Giudice:2007fh}, 
or models of extra
dimensions~\cite{Randall:1999ee,Davoudiasl:1999tf,Csaki:2003dt},
predict diboson resonances at high masses. 
 
%This article describes a search for resonant $WZ$ production in the fully leptonic 
This Letter presents a search for resonant $WZ$ production in the fully leptonic 
decay channels $WZ \rightarrow \ell\nu\ell'\ell'$ ($\ell, \ell'=e, \mu$) using 
\lumi~fb$^{-1}$ of $pp$ collision data collected by the ATLAS detector
at a centre-of-mass energy of $\sqrt{s}=8$ \TeV. 
Four possible leptonic decay channels ($e{\nu}ee$, $e\nu\mu\mu$, $\mu{\nu}ee$ and $\mu\nu\mu\mu$) are
considered.
% and the combination of these four channels is
%referred to as ``All channels combined''. 
To interpret the results, the extended gauge model (EGM)~\cite{Altarelli:1989ff} with a spin-1
%$W^\prime$~boson shall be used as a benchmark signal hypothesis. % to interpret the results obtained. 
$W^\prime$~boson is used as a benchmark signal hypothesis. 
In this model, the couplings of the EGM $W'$ boson to the SM particles 
are identical to those of the $W$ boson, except for its coupling to
$WZ$, which is suppressed with respect to 
the SM $WWZ$ triple gauge coupling by a factor of
$(m_W/m_{W^\prime})^2$ and entails a linear relationship between the resonance width and mass.
%and yields resonance decay widths that scale linearly with mass. 
%Although strong bounds exist on the mass of the $W^\prime$ boson from leptonic
%searches~\cite{Aad:2012dm,ATLAS:2013jma,CMS:2013rca,CMS:2013qca},
%The decay to a pair of gauge bosons 
%is treated as an independent process to possible couplings to fermions, and 
%can be dominant in models with
%leptophobic~\cite{Babu:1996vt,Rizzo:1998ut,Hewett:2011nb} $W^\prime$
%bosons. 
In other scenarios, such as for leptophobic $W'$
bosons~\cite{Babu:1996vt,Rizzo:1998ut,Hewett:2011nb}, the decay to a
pair of gauge bosons can be a dominant channel. 
A narrow $W'$ resonance is predicted in the EGM, with an intrinsic
decay width that is negligible 
with respect to the experimental resolutions on the reconstructed $WZ$ invariant mass. 
%Given the focus on narrow-width signal hypotheses, 
Possible interferences between signal and SM backgrounds are assumed to be small and are neglected. 
%which have Gaussian widths of the order of 25 \GeV for $m(W^{\prime}) =
%200$~\GeV, $100$ \GeV for $m(W') = 1$~\TeV, and $180$ \GeV for a $W'$ boson with
%$m_{W'}=2$~\TeV. %Hence, any spin-1 narrow-width resonance can be
%used to reinterpreat the final results given the signal efficiencies and
%acceptances. 
%Hence, the final results presented here can be reinterpreted in the
%context of any narrow spin-1 resonance for a given signal
%efficiency and acceptance. 
Under these assumptions, the final results presented here can be reinterpreted in terms of any narrow spin-1 resonance for a given signal
efficiency and acceptance. 
%In this context, a phenomenological Lagrangian for heavy vector triplets (HVT)~\cite{Pappadopulo:2014qza}, which can be used

%A phenomenological Lagrangian for heavy vector triplets (HVT)~\cite{Pappadopulo:2014qza}, which can be used
%to reinterpret the results in terms of a large class of %specific 
%models, parameterizes their couplings to fermions and to gauge bosons.
A phenomenological Lagrangian for heavy vector triplets (HVT)~\cite{Pappadopulo:2014qza} has recently been introduced, where the couplings of the new fields 
to fermions and gauge bosons are defined in terms of parameters. By scanning these parameters the generic Lagrangian describes a large class of models.
The triplet field, which mixes with the SM gauge bosons, couples to the fermionic current through the combination of parameters $g^2c_F/g_V$ 
and to the Higgs and vector bosons through $g_Vc_H$, where $g$ is the $SU(2)_{\rm L}$ gauge coupling, the parameter  $g_V$ represents the
coupling strength to vector bosons, and $c_F$ and $c_H$ allow to modify the couplings and are expected to be close to unity in most specific models.
Two benchmark models, provided in Ref.~\cite{Pappadopulo:2014qza},
are used here as well.
In Model A, weakly coupled vector resonances arise from an extension
of the SM gauge group~\cite{Barger:1980ix}. In Model B, the heavy vector triplet  is
produced in a strongly coupled %by a strongly
%Model A, inspired by an extended gauge model with an additional $SU(2)$.
%An additional bi-doublet scalar field obtains a vacuum expectation value, breaking the $SU(2)_1 \times SU(2)_2$
%gauge symmetry to the SM $SU(2)_L$ resulting in a heavy vector triplet which mixes with the SM gauge bosons.
%Model B, where the heavy vector triplet is produced by a strongly 
%coupled
scenario, for example in a Composite Higgs model~\cite{Contino:2011np}.
%where the Higgs results from an $SO(5)/SO(4)$ global symmetry~\cite{Contino:2011np},
%with a vector $\rho$ isotriplet mixing with the SM vector bosons. 
%A parameter $g_V$ is introduced to modify the couplings of fermions and gauge bosons to the heavy vector triplet.
In Model A, the branching fractions to fermions and gauge bosons are comparable, whereas for Model B, fermionic couplings are suppressed.

%Hitherto, 
Direct searches for $WZ$ resonances have been reported by several experiments. 
%The D0 collaboration updated a search for a
%$WZ$~resonance using three diboson decay channels:
%$\ell\nu\ell'\ell'$, $\ell\nu jj$~and $jj\ell\ell$. This search
%excludes an EGM $W^\prime$~boson with mass between 180 and 690 \GeV~\cite{Abazov:2009eu}. The CDF
%collaboration searched for a $WZ$~resonance using the $e\nu jj$~decay channel only and resulted 
%in a lower mass limit of 515~\GeV~\cite{Aaltonen:2010ws}.
The ATLAS collaboration reported on searches for a $W^{\prime}$
resonance using approximately 1~fb$^{-1}$ of data for the
$\ell\nu\ell'\ell'$ channel and 4.7~fb$^{-1}$ of data for the
$\ell\nu jj$ channel, where $j$ is a hadronic jet, both at $\sqrt{s} =7$~\TeV, and excluded an EGM $W^\prime$ boson with mass below 
0.76~\TeV~\cite{ATLAS:lvll_2011} and 0.95~\TeV~\cite{ATLAS:WZlvjj}
respectively.
The advantage of the three-lepton $WZ$ final state over
its partial or fully hadronic final state counterparts is its better sensitivity at 
%the lower end of the $m(WZ)$ mass spectrum due to its significantly
the lower end of the mass spectrum due to its significantly
smaller SM backgrounds and superior mass resolution. 
%The ATLAS collaboration reported on searches for a $W^\prime$ resonance
%using 13 fb$^{-1}$ of data at $\sqrt{s}=8$ \TeV for the $\ell\nu\ell'\ell'$ channel and 4.7 fb$^{-1}$ 
%of data at $\sqrt{s}=7$ \TeV for the $\ell\nu jj$ channel and excluded a EGM $W^\prime$ boson with mass below 
%1180 \GeV~\cite{ATLAS:2013lma} and 950 \GeV~\cite{ATLAS:WZlvjj} respectively. 
The CMS collaboration analysed 5 fb$^{-1}$ of data at $\sqrt{s}=7$ \TeV~
in the $\ell\nu\ell'\ell'$ channel, and 
%EGM $W^\prime$ bosons with masses below 1143 \GeV~\cite{CMS:WZ7\TeV} were excluded. 
EGM $W^\prime$ bosons with masses below 1.143~\TeV~\cite{CMS:lvll_2011}
were excluded. 

%The CMS collaboration has put forth results of their
%search~\cite{CMS:lvll_2011} using the full 2011 data set at $\sqrt{s}
%= 7$~\TeV. With $WZ$ decay final states to electrons and muons, they
%have been able to exclude masses below 1143~\GeV for the EGM
%$W^{\prime}$ boson. 
%The CMS collaboration has recently put forth preliminary results of their 
%search~\cite{CMS:2013vda} using the full 2012 data set. With electron and muon 
%inal states combined and 19.6 fb$^{-1}$ of data used, they have been able to 
%exclude masses between 170 and 1450~\GeV for the EGM $W^\prime$ boson. 
%In the fully hadronic decay channel,
%using boosted vector boson reconstruction technique, CMS excludes the EGM $W^{\prime}$
%below a mass of 1.73 \TeV~\cite{CMS-PAS-EXO-12-024}.

%The search presented here shall use the invariant mass of the
%reconstructed $WZ\rightarrow \ell\nu\ell\ell$ system as its
%discriminating hypothesis-testing variable, thereby requiring a
%handle on the longitudinal component of missing energy. After
%selecting events with three charged leptons and large \etmiss, the missing
%$p_z$ information is derived assuming that the final 3$^{\textrm{rd}}$
%lepton (not the leptons forming the $Z$) combined with the missing
%transverse energy have an invariant mass equal to $m(W) =
%80.4$~\GeV. The resulting quadratic equation has two solutions, and for
%simplicity, the smallest one is kept for if the solution is
%real, and only the real part is kept if the solution is imaginary. 

\section{The ATLAS detector}%, data set and Monte-Carlo modelling}
\label{detector}

The ATLAS detector~\cite{ATLAS1} consists of an inner tracking detector (ID), 
%immersed in a $2$ T magnetic field generated by a superconducting solenoid, 
%electromagnetic (EM) and hadronic sampling calorimeters, 
electromagnetic (EM) and hadronic calorimeters, 
and a muon spectrometer.  
%utilizing the eight air-core toroidal magnets with a maximum strength of $4$ T. 
The ID is immersed in a 2 T axial magnetic field, generated by a superconducting solenoid, and 
consists of a silicon pixel detector, a silicon microstrip detector, and a transition radiation tracker. 
The ID provides a pseudorapidity coverage of $|\eta| < 2.5$.\footnote{ATLAS uses a right-handed coordinate system 
with its origin at the nominal interaction point (IP) in the centre of the detector 
and the $z$-axis along the beam pipe. 
The $x$-axis points from the IP to the centre of the LHC ring, 
and the $y$-axis points upward. 
Cylindrical coordinates $(r,\phi)$ are used in the transverse plane, 
$\phi$ being the azimuthal angle around the beam pipe. 
The pseudorapidity is defined in terms of the polar angle $\theta$ as $\eta=-\ln\tan(\theta/2)$.
The separation between final-state particles is defined as $\Delta R = \sqrt{(\Delta\eta)^2 + (\Delta\phi)^2}$.
The transverse momentum is denoted by $p_{\rm T}$.}
%~\footnote{ATLAS uses a right-handed coordinate system 
%with its origin at the nominal $pp$ interaction point (IP) in the centre of the detector 
%and the $z$-axis along the beam. 
%The $x$-axis points from the IP to the centre of the LHC ring, 
%and the $y$-axis points upward. 
%Cylindrical coordinates ($r, \phi$) are used in the transverse plane, 
%$\phi$ being the azimuthal angle around the $z$-axis. 
%The pseudorapidity is defined in terms of the polar angle $\theta$ as $\eta = -\ln \tan (\theta/2)$. 
%The separation between final state particles is defined as $\Delta R = \sqrt{(\Delta\eta)^2 + (\Delta\phi)^2}$. 
%The transverse momentum is denoted as $p_{\rm T}$.}.
%

%\cite{coordinate}.  
%inside a transition radiation tracker in the region $|\eta| < 2.0$. 
The EM calorimeters are composed of interspersed lead and liquid argon, acting as absorber and active material respectively, 
with high granularity in both the barrel ($|\eta| < 1.475$) and end-cap up to the end of the tracker acceptance ($1.375 < |\eta| < 2.5$),
and somewhat coarser granularity from $|\eta|=2.5$ to 3.2. 
%($1.375 < |\eta| < 3.2$) regions. 
%In the forward regions ($3.15<|\eta|<4.9$) where radiation is
%intense, the calorimeters have lower granularity and
%sampling fraction, as well as absorbers that are made of copper and tungsten instead
%of lead. 
%In the radiation intense region ($3.15 < |\eta| < 4.9$), the forward calorimeters have relatively smaller 
%sampling cells and replace lead by a combination of copper and tungsten absorbers. 
The hadronic calorimeter uses steel and scintillator tiles in the barrel region, 
while the endcaps use liquid argon as the active material and 
copper as an absorber. %tile scintillating ($|\eta| < 1.7$) hadronic calorimeter, having an approximate 
%projective geometry (equal in $\eta$ and $\phi$), utilizes steel as its absorber, whereas the end-cap 
%liquid-argon hadronic calorimeters ($1.5 < |\eta| < 3.2$) utilize copper. 
%The EM and hadronic calorimeters are important for the missing transverse energy (\etmiss) measurement.
The muon spectrometer (MS) is based on three large superconducting air-core toroids arranged with an eight-fold 
azimuthal coil symmetry around the calorimeters. Three layers of
precision tracking chambers, consisting of drift tubes and cathode strip chambers, 
enable precise muon track measurements in the pseudorapidity range of $|\eta| < 2.7$, and resistive-plate 
and thin-gap chambers provide muon triggering capability in the range of $|\eta| < 2.4$.

\section{Data and Monte Carlo~modelling} \label{data}

The data analysed here were collected by the ATLAS detector at the LHC in $pp$ collisions at
% a centre-of-mass energy of 
$\sqrt{s}=8$ \TeV~during the 2012 data-taking run. 
%Events are selected using unprescaled single-isolated electron and muon triggers with a 24 \GeV~transverse momentum threshold, 
%or unprescaled single electron and muon triggers without isolation criteria with 60 \GeV~and 36 \GeV~transverse momentum thresholds respectively.  
Events are selected using a combination (logical OR) of isolated and non-isolated single-lepton ($e$ or $\mu$) triggers. The $p_{\rm T}$ thresholds are 24 \GeV~ for isolated single-lepton triggers and 60 (36) \GeV~for non-isolated single-$e$ ($\mu$) triggers.
%Given the requirement for three high $p_{\rm T}$ leptons in the final 
%state, the trigger efficiency is above $99.5$\%.
%The requirement of three high-$p_{\rm T}$ leptons in the final state gives a trigger efficiency above 99.5\%. 
The requirement that three high-$p_{\rm T}$ leptons are in the final state gives a trigger efficiency above 99.5\%.
After data-quality requirements are applied, the total integrated luminosity is \lumi~fb$^{-1}$ with an uncertainty of $2.8$\%~\cite{Lumi}.

The baseline EGM $W'$ signals are generated with \pythia~8.162~\cite{pythia}
and the MSTW2008LO~\cite{mstw2008lo} parton distribution function
(PDF) set. 
The production cross section times branching fraction (with 
%$W \rightarrow e\nu_{e}, \mu\nu_{\mu}, \tau\nu_{\tau}$, where all $\tau$ decays are considered, and $Z \rightarrow ee, \mu\mu$) are scaled to their theoretical 
$W \rightarrow e\nu, \mu\nu, \tau\nu$, where all $\tau$ decays are considered, and $Z \rightarrow ee, \mu\mu$) are scaled to their theoretical 
predictions at next-to-next-to-leading order (NNLO) using
ZWPROD~\cite{ZWPROD}, which are $1.43$~pb for $m_{W^{\prime}} =
200$~\GeV, $4.12$~fb for $m_{W^{\prime}}= 1$~\TeV, and $0.08$~fb for $m_{W^{\prime}} = 2$~\TeV.
%Regarding the $W \rightarrow \tau \nu_{\tau}$ component, only the
In the $W \rightarrow \tau \nu$ component, only the
leptonic $\tau$ decays enter the signal acceptance, albeit slightly
and only at high signal mass, whereas the $Z\rightarrow \tau \tau$ component is totally negligible. 
The intrinsic decay widths 
%$\Gamma$ 
of the EGM $W^{\prime}$ scale
linearly with $m_{W^{\prime}}$ at high mass and are 
$5.5$~\GeV~for $m_{W^{\prime}} = 200$~\GeV, $36$~\GeV~for $m_{W^{\prime}} = 1$~\TeV,
and $72$~\GeV~for $m_{W^{\prime}}= 2$~\TeV. 
These are significantly less than the experimental resolutions, which
have Gaussian widths of the order of $25$~\GeV~for $m_{W^{\prime}} =
200$~\GeV, $100$~\GeV~for $m_{W^{\prime}} = 1$~\TeV, and $180$~\GeV~for $m_{W^{\prime}}=2$~\TeV. 
% to interpret the final results. 
%A mass dependent scaling at parton level was found to affect negligibly the signal kinematical distributions, and was hence neglected. 
MC samples were produced for the EGM $W'$ signal from 200 \GeV~to 400 \GeV~at intervals of 50 \GeV~and from 400 \GeV~to 2 \TeV~at intervals of 200 \GeV. 
An interpolation procedure is adopted to obtain the distributions for mass points between 200 \GeV~and 400 \GeV~with 25 \GeV~step size
and from 400 \GeV~to 2 \TeV~with 50 \GeV~step size. 
\renewcommand{\arraystretch}{1.2}
\begin{table}[htbp]
\caption{Overview of the primary MC samples. The backgrounds
from misidentified jets are estimated from the data.}\vspace*{1mm}
\label{mc}
%\resizebox{\columnwidth}{!}{
\centering
\begin{tabular}{c|c|c|c}
\hline
\hline
Process & Generator & Parton Shower & PDF \\
\hline
$W^{\prime}$ & \pythia & \pythia & MSTW2008LO\\
\hline
%$WZ$ & \powheg & \pythia & \multirow{3}{*}{{CT10}{~\cite{ct10}}} \\
$WZ$ & \powheg & \pythia & \\
$ZZ$ & \powheg & \pythia & CT10 \\
$Z\gamma$ & \sherpa & \sherpa & \\
%$W\gamma$ & \alpgen & 
%herwig+\jimmy & 
\hline
$t\bar{t}$+$W/Z$ & \madgraph & \pythia & CTEQ6L1 \\
\hline
\hline
\end{tabular}%}
\end{table}

%The dominant irreducible SM $WZ$ background is modelled by \powheg~\cite{powheg}, a next-to-leading-order (NLO) event generator 
The dominant SM $WZ$ background is modelled by \powheg~\cite{powheg}, a next-to-leading-order (NLO) event generator 
combined with the NLO CT10 PDF set~\cite{ct10}. % containing NLO matrix elements while
Background events arising from $ZZ$ are modelled with \powheg, while those from $t\bar{t}+W/Z$ processes 
are generated with \madgraph~5.1.4.8~\cite{madgraph} together with the
CTEQ6L1~\cite{cteq} PDF set. All these events are interfaced with \pythia, using the AU2 tune~\cite{pythia_tune} for parton showering.

A second category of background arises from photons misidentified as electrons, mainly from $Z\gamma$ production.
A photon can be misreconstructed as an electron if it lies close to 
a charged particle track or if the photon converts to $e^+ e^-$ after interacting with the material 
in front of the calorimeter. This contribution is estimated using simulated $Z\gamma$ MC events 
generated with \sherpa~1.4.0~\cite{sherpa}. 

Finally, a third category of background includes all other sources
where one or more jets are misidentified as an isolated lepton. 
The contributions from these {\it fake} backgrounds
are estimated using a data-driven method as described in
Section~\ref{background}. 
The contribution from events with only one jet misidentified as an isolated lepton is found to be 
dominant while those with more than one are found to
be negligible. Thus, in this analysis the {\it fake} backgrounds are denoted by $\ell\ell'+$jets. 

An overview of the major MC samples used is presented in Table~\ref{mc}. 

Monte Carlo (MC) events are processed through the full detector
simulation~\cite{infrastructure} using {\sc geant4}~\cite{geant}, and their reconstruction is performed with the same software used to reconstruct data events.
Correction factors for lepton reconstruction and identification efficiencies are applied to the simulation to account for differences with respect to data. 
The simulated lepton four-momenta are tuned, via calorimeter energy
scaling and momentum resolution smearing, to reproduce the distributions
%observed in data from leptonic $W$, $Z$ and $J/\psi$ decays.
observed in data from leptonic $W$, $Z$ and $J/\psi$ decays after calibration.
Furthermore, additional inelastic $pp$ collision events are overlaid
with the hard scattering process in the MC simulation and then reweighted to reproduce 
the observed average number of interactions per bunch-crossing in data. 

\section{Object reconstruction} \label{selections}

Electron candidates are reconstructed in the region of the EM
calorimeter with $|\eta| < 2.47$
by matching the calorimeter clusters to the tracks in the ID. 
The transition region between the barrel 
and endcap calorimeters ($1.37 < |\eta| < 1.52$) is excluded. 
%The calorimeter clustering and ID track reconstruction must
%satisfy a set of criteria expected from electron energy deposition. 
Candidate electrons must satisfy the \texttt{medium} 
quality definition~\cite{ATLAS:EMID} re-optimized for 2012 data-taking conditions, which is 
based on a set of requirements on the calorimeter shower shape, track quality, and track 
matching with the calorimeter cluster. The longitudinal impact parameter $z_0$ of the associated 
%track with respect to the PV must satisfy $|z_0 \sin \theta| <
track with respect to the primary vertex (PV), which is
defined as the vertex with the largest sum of squared transverse momenta of associated tracks,
must satisfy $|z_0 \sin \theta| < 0.5$~mm. 
The transverse impact parameter $d_0$ of the associated track 
must satisfy $|d_0/\sigma_{d_0}| < 6$, where $\sigma_{d_0}$ is the uncertainty on
the measurement of $d_0$. 
To reduce the background due to jets misidentified as electrons,
electron candidates are required to be isolated in both the calorimeter and the ID. 
The isolation requirements are $R^{\textrm{iso}}_{\textrm{Cal}}<0.16$ and
$R^{\textrm{iso}}_{\textrm{ID}}<0.16$, where $R^{\textrm{iso}}_{\textrm{Cal}}$ is the total transverse energy 
recorded in the calorimeters within a cone of size
%$\Delta R = \sqrt{(\Delta \eta)^2 + (\Delta \phi)^2}=0.3$ 
$\Delta R = 0.3$ around the lepton direction,
excluding the energy of the lepton itself, divided by the lepton
$E_{\rm T}$, and $R^{\textrm{iso}}_{\textrm{ID}}$ is the sum of the $p_{\rm T}$ of the tracks in a cone of size $\Delta R=0.3$ around 
the lepton direction, excluding the track of the lepton, divided by the lepton $p_{\rm T}$. 

Muon candidates are reconstructed within the range $|\eta| < 2.5$ 
by combining tracks in the ID and the MS. 
%A robust reconstruction is ensured by requiring %a minimum number of
% silicon microstrip and pixel hits associated with the track. 
%The ID tracks associated with muons that are identified inside the ID acceptance are required
%to have a minimum number of associated hits in each of the ID sub-detectors to ensure robust track reconstruction.
%Robust reconstruction is ensured by requiring a minimum number of 
%silicon microstrip and pixel hits to be associated with the ID tracks. 
Robust reconstruction is ensured by requiring a minimum number of hits in each of the sub-detectors of ID to be associated with the reconstructed ID tracks.
Moreover, the muon reconstructed track must satisfy
the requirements $|z_0 \sin \theta| < 0.5$~mm and $|d_0/\sigma_{d_0}| < 3.5$. 
%Electrons have a worse impact parameter resolution than muons due to bremsstrahlung. 
The measured momenta in the ID and the MS are required to be consistent with each other by 
satisfying $|(q/p)^{\textrm{ID}} - (q/p)^{\textrm{MS}} | < 5 \sigma$, where 
$(q/p)^{\textrm{ID}}$ and $(q/p)^{\textrm{MS}}$ are the charge $q$
over momentum $p$ in the ID and the MS respectively, and 
$\sigma$ is the total uncertainty on the difference between $q/p$
measurements in the ID and the MS. 
%over momentum $p$ in the ID and MS respectively, and where $\sigma$ is the combined 
%uncertainty for $q/p$ measurements in the ID and MS. 
The muon isolation requirements are $R^{\textrm{iso}}_{\textrm{Cal}}<0.2$ 
and $R^{\textrm{iso}}_{\textrm{ID}}<0.15$. 

When the $Z$ boson has high momentum ($\gtrsim 600$~\GeV), its collimated lepton decay
products can be within a cone of size $\Delta R = 0.3$. 
To maintain a high efficiency for high-mass signals the isolation requirements imposed on the leptons are modified
%The isolation requirements
%imposed on the leptons are therefore modified 
to not include in the calculation of $R^{\textrm{iso}}_{\textrm{Cal}}$ and
$R^{\textrm{iso}}_{\textrm{ID}}$ the energy and momenta of any
%proximate same-flavour lepton. This enables to maintain a high signal
close-by same-flavour leptons. 
%This enables to maintain a high 
%efficiency for high mass signals. 
For a $m_{W^{\prime}}
= 1.4$~\TeV~signal, the relative efficiency gain, with respect to the
selection without modifying the isolation requirements, is of the
order of 60\%. Finally, to reduce photon conversion backgrounds from muon radiation, if a muon and an electron are separated by less than
$\Delta R=0.1$ from each other, the electron candidate is
discarded. 

The missing transverse momentum, with magnitude \etmiss, is the momentum imbalance in the transverse plane. The \etmiss~is calculated from the negative vector sum of 
%The missing transverse momentum (\etmiss) is the magnitude of the 
%vector momentum imbalance in the transverse plane.
%momentum imbalance in the transverse plane.
%The \etmiss~is calculated as the negative vector sum of 
the transverse momenta of all reconstructed objects, 
including muons, electrons, photons and jets, as well as clusters of calorimeter cells not associated with these objects~\cite{etmiss_paper}. 

Attributing the \etmiss~to the transverse component of the neutrino
momentum, its longitudinal component ($p^{\nu}_{\rm z}$) is derived by 
%assuming that the lepton, from the $W$ boson decay, and 
requiring that the neutrino and the lepton attributed to the $W$ boson decay 
have an invariant mass equal to the pole mass of
the $W$ boson: 80.385 \GeV~\cite{Wmass}. 
This constraint results in a quadratic equation with two solutions for $p^{\nu}_{\rm z}$.
If the solutions are real the one with the smaller absolute value is kept. If the solutions are complex only the real part is kept. 
%The one with the smaller absolute value is kept if the solutions are
%real, and only the real part is kept if the solutions are complex
%numbers. 
In general, about 30\% of the events are found to have complex solutions, 
mainly due to the~\etmiss~resolution at the reconstruction level. 
%The fraction decreases for the high mass signals ($\gtrsim 600$ \GeV) to 20\%
%as the $W$ boson becomes more on-shell and the resolution for electron and jet improves as the energy increases.
The invariant mass of the  $WZ\rightarrow \ell\nu\ell'\ell'$ system is 
reconstructed from the four-vectors of the candidate $W$ and $Z$ bosons
%defined as  
%$m_{WZ} = \sqrt{(E_{\ell\nu} + E_{\ell'\ell'})^2 - (\vec{p}_{\ell\nu} + \vec{p}_{\ell'\ell'})^2}$, 
%where $E_{\ell\nu}$ and $E_{\ell'\ell'}$ are the total energy of the $\ell\nu$ and $\ell'\ell'$ systems respectively, 
%$\vec{p}_{\ell\nu}$ and $\vec{p}_{\ell'\ell'}$ are the total momentum of the $\ell\nu$ and $\ell'\ell'$ systems respectively.   
and is used as the discriminating variable for the signal.%signal discriminant variable. %discriminating hypothesis-testing variable.

\section{Event selection}

%Candidate events are required to contain at least one primary vertex (PV), 
%which is defined as the vertex with the largest sum of squared transverse momenta of associated tracks. 
%This PV must also have at least 
%three associated tracks with $p_{\rm T}>0.4$ \GeV. 

The PV of the event must have at least 
three associated tracks with $p_{\rm T}>0.4$ \GeV.
Candidate $WZ \rightarrow \ell\nu\ell'\ell'$ events are then required to have exactly 
three charged leptons with $p_{\rm T} > 25$~\GeV~and \etmiss$>25$ \GeV. 
Events are rejected if a fourth lepton is found with $p_{\rm T} > 20$~\GeV. 
At least one of the three leptons is required to be geometrically matched 
%to a lepton reconstructed by the trigger.
to an object that fired the trigger.
Two opposite-sign same-flavour leptons are required to have an invariant 
mass ($m_{\ell\ell}$) within 20 \GeV~of the $Z$ boson pole mass: 91.1875 \GeV~\cite{ALEPH:2005ab}. 
If two possibilities exist, the pair that has $m_{\ell\ell}$ closest to 
the $Z$ boson pole mass is chosen to form the $Z$ candidate.
%To reduce the contributions from $Z+$jet events, the candidate
To suppress the $Z+$jets background where one jet is reconstructed as an
isolated electron, the electrons used in the reconstruction of the $W$
bosons are required to satisfy tighter identification criteria (\texttt{tight})
than those required for the leptons used in the reconstruction of $Z$
boson decays (\texttt{medium}). These stricter criteria are described in Ref.~\cite{ATLAS:EMID}.

To improve the sensitivity to resonant signals, events are further required to have $\Delta y (W,Z) < 1.5$, where $\Delta y (W,Z)$ is the
rapidity\footnote{Rapidity is defined as $y = (1/2)\ln[(E+p_{\rm z})/(E-p_{\rm z})]$.}
 difference between the $W$ and $Z$ bosons. This selection has 
%an efficiency beyond 82\% for all $W'$ masses and reaching 94\% for
an efficiency exceeding 82\% for all $W'$ masses and reaching 94\% for
$m_{W^{\prime}} = 200$~\GeV. 

%To improve the sensitivity of the search to the EGM $W'$ boson, 
Finally, two signal regions are defined, one more sensitive for high-mass $W'$ signals ($m_{W^{\prime}} \gtrsim 250$ \GeV) and the 
other one for low-mass $W'$ signals ($m_{W^{\prime}} \lesssim 250$ \GeV). 
The high-mass signal region (SR$_\textrm{HM}$) is defined by the additional
requirement
%s of $\Delta y (W,Z) < 1.5$ and 
$\Delta \phi(\ell, E^{\textrm{miss}}_{\rm T}) < 1.5$, 
where 
%$\Delta y (W,Z)$ is the rapidity difference between the $W$ and $Z$ bosons, and 
$\Delta \phi(\ell, E^{\textrm{miss}}_{\rm T})$ is the
azimuthal angle between the lepton attibuted to the $W$ candidate decay and the missing
transverse momentum vector. %However, the selection upon $\Delta y (W,Z)$
%is inefficient for signal mass hypotheses below $leq 350$~\GeV. Therefore, a 
Conversely, the low-mass signal region (SR$_\textrm{LM}$) is required to have $\Delta \phi(\ell, E^{\textrm{miss}}_{\rm T})>1.5$, 
%where low mass signal
%efficiency is recovered. 
which has high acceptance for low-mass signals.

%The EGM $W'$ signal samples are generated with $m_{W'}$ from 200 \GeV to 
%2 \TeV with a step size of 200 \GeV. 
%For the EGM $W'$ signals, 
%the acceptance times efficiency ($\mathcal{A} \times \epsilon$) in \srhm increases from 2\% to 30\% for $m_{W'}$ from 200 \GeV to 2 \TeV, 
%while the $\mathcal{A} \times \epsilon$ in \srlm decreases from 8\% to 0\%. 

\section{Background estimations}  \label{background}

\begin{figure}[htbv]
\includegraphics[width=\columnwidth]{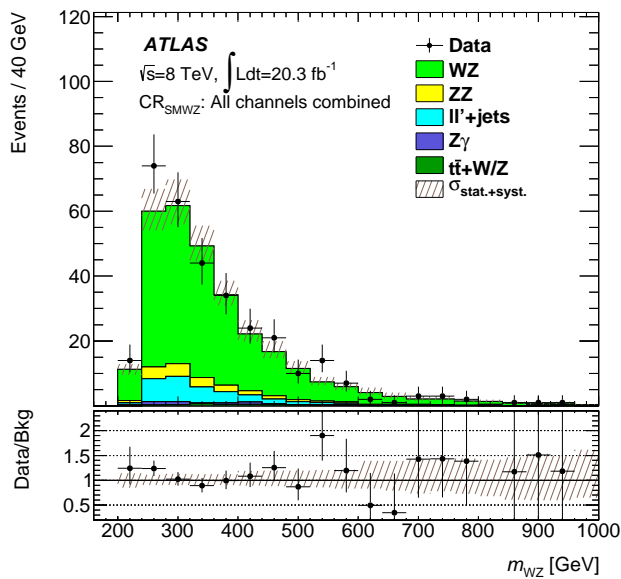}
\caption{Distribution of $WZ$ invariant mass ($m_{WZ}$) in the SM $WZ$ control region (CR$_{\textrm{SMWZ}}$)
  for the four $\ell\nu\ell'\ell'$ channels combined. The uncertainty bands upon the expected background include
both the statistical and systematic uncertainties in the MC simulation and the fake-background estimation added in quadrature.} \label{wz_control}
\end{figure}

%The dominant irreducible backgrounds come from the combination of the SM $WZ$, $ZZ$ and $t\bar{t}+W/Z$ processes with three prompt leptons in the final state. 
%The major irreducible backgrounds come from the SM $WZ$, $ZZ$ and $t\bar{t}+W/Z$ processes with three prompt leptons in the final state. 
The major backgrounds come from the SM $WZ$, $ZZ$ and $t\bar{t}+W/Z$ processes with at least three prompt leptons in the final state. 
%The contributions from these backgrounds are estimated using the MC samples.
%Background due to misidentification of a photon as an electron , mainly coming from the $Z\gamma$ process, is also estimated using MC.
A control region dominated by SM $WZ$ events (CR$_{\textrm{SMWZ}}$) is defined to check the modelling of the MC predictions for these backgrounds. 
The selection criteria used for this region are similar to those for the signal regions except that the requirement on $\Delta y(W, Z)$ is 
reversed and the requirement on $\Delta \phi(\ell, E^{\textrm{miss}}_{\rm T})$ is removed. 
The reversal of the $\Delta y(W,Z)$ selection reduces possible signal contamination to negligible levels, assuming previous exclusion results~\cite{ATLAS:lvll_2011,CMS:lvll_2011}. 
In total, 323 events are observed in data for all four channels
combined and the SM backgrounds are expected to be 
$298 \pm 4 {\textrm{(stat.)}} \pm 26{\textrm{(syst.)}}$ 
events, where the computation of the systematic uncertainties is
detailed in Section~\ref{syst}. 
%Good agreement is found between
%observed data and SM predictions also in terms of the shape of various
%kinematical distributions. 
Good agreement is also found between data and the SM predictions in the shapes of various kinematical distributions.
The $m_{WZ}$ distribution in the SM $WZ$ control region is shown in Fig.~\ref{wz_control}.
%Fig.~\ref{wz_control} in the SM $WZ$ control region (CR$_{\textrm{SMWZ}}$). 

Contributions from the $\ell\ell'$+jets background,
%where at least one lepton originates from hadron decays in jets, are estimated using a data-driven method. 
where at least one lepton originates from hadronic jets, are estimated using a data-driven method. 
%A lepton-like jet is defined as a jet that satisfies all lepton selection criteria but, in the muon case, fails either the calorimeter or track isolation requirement, 
A lepton-like jet is defined as a jet that is reconstructed as a lepton and satisfies all lepton selection criteria but, in the muon case, fails either the calorimeter or track isolation requirement, 
or, in the electron case, fails the isolation or \texttt{medium} quality requirement but passes a looser set of electron identification quality requirements. 
A ``fake factor'', defined as the number of events in which a
jet satisfies the nominal lepton selection criteria divided by the number of events in which a jet
satisfies the lepton-like jet criteria, is computed. It can be
interpreted as the probability that a lepton-like jet 
%passes the
is instead reconstructed as a
nominal lepton. The fake background is dominated by events
with one jet misidentified as an isolated lepton, while contributions
from other processes with two or three jets misidentified as isolated
leptons are found to be negligible. 
The fake background is thus estimated by applying the fake factor to a data sample (denoted as ``tight+loose sample'') selected using all signal 
selection criteria except for a requirement that one of the three leptons must be a lepton-like jet. 
Since the electron identification and isolation requirements are different for
those coming from a $Z$ or a $W$ candidate decay, the electron fake factor
is calculated separately for these two cases. 

The lepton fake factor is measured in two different data samples: dijet and $Z+$jets events. 
In both cases the tag-and-probe method~\cite{tagprobeCDF,tagprobeD0} is used, but the tag objects are different. 
The larger number of events within the dijet sample permits a measurement of the
dependence of the lepton fake factor on the lepton $p_{\rm T}$ or $\eta$.
Using the $Z+$jets sample, on the other hand, leads to a measurement where the kinematic distributions and flavour compositions are closer to that of the signal region, albeit with significantly
fewer events allowing only a measurement of the fake factor as a single number.  

In the tight+loose sample and the two samples used for the fake-factor measurement, 
the backgrounds containing prompt leptons are estimated using MC simulation and subtracted from the data samples.
These include the production of $Z+$jets simulated with \alpgen~2.14~\cite{Mangano:2002ea}, 
$t\bar{t}$ with \mcatnlo~4.03~\cite{Frixione:2002ik}, $W+$jets and $W\gamma$ with
\alpgen, as well as the previously mentioned $WZ$, $ZZ$, $Z\gamma$,
and $t\bar{t}+W/Z$ MC samples. % and QCD multijets events with \pythia. 
The parton showering is modelled by
\herwig/\jimmy~\cite{b-herwig,jimmy} for $Z+$jets, $t\bar{t}$,
$W+$jets, and $W\gamma$ events. 
The events remaining after subtraction are thus the expected lepton yields due to misidentified jets.  

The dijet sample is selected with one tag jet 
and one probe jet that are almost back-to-back, with $\Delta\phi > 2.5$. 
The tag jets are normal hadronic jets and the probe jet is required to satisfy the
%pass the lepton-like jet or the nominal lepton selection criteria.
selection criteria for a lepton-like jet or a nominal lepton.
The tag jets are reconstructed up to $|\eta|=4.5$ from calorimeter clusters with the anti-$k_t$
algorithm~\cite{antikt} using a distance parameter of 0.4 and are calibrated to the hadronic energy scale.
They are required to have $p_{\rm T} > 25$~\GeV.
For jets with $p_{\rm T} < 50$ \GeV~and $|\eta| < 2.4$, the
%scalar $p_{\rm T}$ sum of the tracks falling into the jet area consistent
%with the primary vertex must equate to at least 50\% of the total
scalar $p_{\rm T}$ sum of the tracks that are associated with the PV and that fall into the jet area 
must be at least 50\% of the 
scalar $p_{\rm T}$ sum of all tracks falling into the same jet area. 
The dijet events are selected by single-muon and single-photon triggers, 
with $p_{\rm T}$ and $E_{\rm T}$ thresholds of 24 and 20~\GeV~in the muon
and electron cases respectively. 
The muon/electron requirements at the trigger level are looser than the lepton-like jet selection criteria in order to allow for an 
unbiased measurement of the lepton fake factor. 
To better mimic the kinematic properties of the signal region, the \etmiss~is required to be higher than 25~\GeV,
which also helps reject the $Z+$jets background. 
%because we expect most misidentified jets to be associated with the
%lepton the $W$ decay. 
%The \etmiss~is required to be higher than 25~\GeV 
%which is the same cut as applied in the signal regions and helps reject the $Z+$jets background.
The probe jet and the missing transverse momentum are required to have a transverse mass smaller than 40 \GeV~to suppress the $W+$jets background. 
The probe jet is then examined to determine 
%the probability that it passes the nominal lepton selection criteria or only those of the lepton-like jet.% requirements. 
whether it satisfies the nominal lepton selection criteria or those of the lepton-like jet.

The $Z+$jets sample is defined as having one same-flavour opposite-charge lepton pair consistent with the $Z$ boson decay as the tagged object, % that is consistent with a 
and 
%a probe jet that passes the lepton-like jet or the nominal lepton selection criteria.
a probe jet that satisfies the selection criteria for a lepton-like jet or a nominal lepton.
They are selected by a set of single-lepton and dilepton triggers to improve the trigger efficiency.
To suppress the contribution from prompt leptons from $WZ$ production, events are required to have $E^{\textrm{miss}}_{\rm T}<25$ \GeV.
%and the transverse mass of the lepton-\etmiss~system $M_{\rm T}^W=\sqrt{2p_{\rm T}^\ell E_{\rm T}^{\textrm{miss}}(1-\cos \Delta \phi)}<25$ \GeV, 
%where $p_{\rm T}^\ell$ is the transverse momentum of the third lepton and $\Delta \phi$ is the opening angle between the third lepton 
%and the \etmiss~direction in the transverse plane.  
%---This cut is used in the Z+jet Control region ---
The probe jet is used for measuring the fake factor.
  
In both the dijet and $Z+$jets samples, several sources of systematic uncertainty for the measurement of the
fake factors are considered,
%Several systematic sources are considered for the fake factor measurements for both samples, 
%stemming from the trigger efficiencies threshold, 
stemming from the trigger bias, 
kinematic and flavour differences with respect to the 
signal region, the \etmiss~threshold requirement, and prompt-lepton subtraction. In the dijet sample, 
possible biases related to the tag-jet $p_{\rm T}$ threshold, the transverse mass requirement on 
the probe jet and \etmiss~system, and the azimuthal angle 
between the tag jet and the probe jet are also
considered. Likewise, additional biases associated with the measurement
in the $Z+$jets sample, such as potential systematic kinematic differences between the low- and high-\etmiss~regions, are also considered.
The total uncertainties on the fake factors measured using the dijet sample 
ranges from 8\% to 33\% for muons with $p_{\rm T}<50$ \GeV~and electrons with $p_{\rm T}<70$ \GeV. 
Beyond the above $p_{\rm T}$ ranges the fake factors are assigned a $\gtrsim$ 100\% systematic uncertainty due to the subtraction of prompt backgrounds. 
The total uncertainties on the fake factors measured using the $Z+$jets sample range from 27\% to 36\% for different lepton flavours and definitions. 
The uncertainties on the fake factors are applied to the fake-background estimate as normalization uncertainties.
%Contamination from prompt $Z$+jet, in the case the misidentified jet is 
%used to reconstruct the $Z$ boson. 

%The fake factors, which are of the order of 0.1, measured in both samples are found to be consistent
The fake factors, which are of the order of 0.1 for both lepton flavours, are measured in both
%samples and are found to be consistent with each other. 
%samples and are found to be consistent with each and give compatible results. 
%samples and give compatible results. 
samples. 
%other within the uncertainties.
%The $p_{\rm T}$-binned central values from the dijet sample measurement
%are the ones used in this analysis, while differences in the fake factors between
%the two samples are used as additional systematic uncertainties and are
%the dominant source of uncertainty of the order of~$\gtrsim60$\%.  
The $p_{\rm T}$-binned central values from the dijet sample measurement 
are the ones used in this analysis. The differences between 
the fake factors from the two samples can be up to $\sim60\%$ 
%$\gtrsim60$\%
and are the dominant contributions to the fake-factor uncertainty.
%Differences between them are used as additional systematic uncertainties, 
%but the estimations from the dijet sample are used from the central
%value. 
%The fraction of events with misidentified jets contributing to the reconstruction of the $Z$ boson
%is estimated to be roughly 20\% (40\%) in \srhm~(\srlm) in relation to
%those that fake the signature of a $W$ boson. 
%a control sample enriched in $\ell \ell'$+jet events with the jet misidentified
%as a lepton is defined.

\begin{figure}[htbv]
\includegraphics[width=\columnwidth]{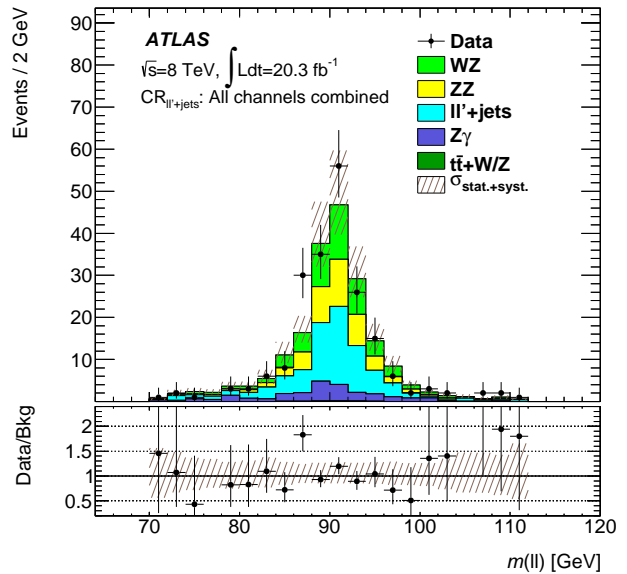}
\caption{$Z$ candidate invariant mass distribution in the $\ell\ell'+$jets background control region (CR$_{\ell\ell'+\textrm{jets}}$). 
The uncertainty bands upon the expected background include
both the statistical and systematic uncertainties in the MC simulation and the fake-background estimation added in quadrature.}
\label{Zjetcontrol}
\end{figure}

The observed and predicted background event yields are compared in
a $\ell \ell'$+jets-enriched control region (CR$_{\ell\ell'+\textrm{jets}}$) where events are required to have the same lepton selection and 
$Z$ mass requirement as in the nominal signal selection but with $E^{\textrm{miss}}_{\rm T}$ less than 25 \GeV~and the transverse mass 
of the $W$ candidate %$W$ lepton-\etmiss~system 
%$m_{\rm T}^W=\sqrt{2p_{\rm T}^\ell E_{\rm T}^{\textrm{miss}}(1-\cos \Delta \phi)}$ 
less than 
25 \GeV.
%where $p_{\rm T}^\ell$ is the transverse momentum of the third
%lepton and $\Delta \phi$ is the opening angle between the third lepton
%and the \etmiss~direction in the transverse plane. a
%
%   to validate this 
%control region to make sure the central values of the fake factors are correct. The control sample is 
%selected by requiring three leptons in the event with a same-flavour lepton pair consistent with the 
%decays from a $Z$ boson. Events are selected to have $E^{\textrm{miss}}_{\rm T} < 25$~\GeV to be orthognal to 
%the signal events. The selected events are mainly due to a jet misidentified as a lepton from the 
%$Z+$jets process. The comparison between data and SM predictions for the invariant mass distribution 
%of the sysem is shown in Fig.~\ref{Zjetcontrol},
%For the $\ell\ell'+$jets control region, in total 204 events are observed in data with a SM expectation of $195.3 \pm$ (stat.) $\pm 38.2$ (syst.) events. 
In this region, a total of 204 events are observed in data with an SM
expectation of 
%$195 \pm$ 4 (stat.) $\pm 38$ (syst.) 
$195 \pm 4 {\textrm{(stat.)}} \pm 38 {\textrm{(syst.)}}$ 
events. Good agreement is found between observed data and estimated background 
for various kinematic distributions. 
%one example is shown in Fig.~\ref{Zjetcontrol} for the $Z$ candidate invariant mass distribution. 
The $Z$ candidate invariant mass distribution is shown in Fig.~\ref{Zjetcontrol}. 

%The estimated fake backgrounds in the signal regions are found to be consistent with each other for each individual 
%channel using fake factors measured in both samples. %, as shown in Tab.~\ref{fake_result}. 

%\renewcommand{\arraystretch}{1.5}
%\begin{table}[htbp]
%\centering  
%\tiny
%\begin{tabular}{c|c|c|c}
%\hline
%\hline
%\specialcell{Fake factor \\ sample} & Channel & \specialcell{Fake background
%  \\ in \srhm} & \specialcell{Fake background \\ in \srlm}\\
%\hline
%\multirow{4}{*}{dijet} & $eee$ & $12.7\pm1.0\pm5.6$ & $13.3\pm1.1\pm5.4$ \\ 
% & $ee\mu$     & $19.3\pm1.7\pm4.1$ & $13.0\pm1.5\pm2.9$ \\
% & $e\mu\mu$   & $14.0\pm1.0\pm6.9$ &$17.0\pm1.3\pm6.2$ \\
% & $\mu\mu\mu$ & $23.1\pm1.9\pm7.2$ & $22.0\pm1.8\pm6.8$ \\
%\hline
%\multirow{4}{*}{$Z+$jets} & $eee$ & $19.7\pm1.5^{+6.5}_{-6.3}$ & $18.4\pm1.5^{+6.2}_{-6.0}$ \\
% & $ee\mu$     & $29.1\pm2.6^{+10.0}_{-9.4}$ & $19.1\pm2.2^{+6.4}_{-6.2}$ \\
% & $e\mu\mu$   & $24.5\pm1.8^{+9.0}_{-8.3}$ & $28.4\pm2.2^{+9.7}_{-10.7}$ \\
% & $\mu\mu\mu$ & $36.1\pm3.0^{+12.9}_{-12.0}$ & $33.0\pm3.0^{+11.0}_{-12.2}$\\
%\hline
%\hline
%\end{tabular}
%\caption{Estimated fake background contributions in \srhm~and \srlm~for each individual decay channel. 
%Results using fake factors measured with dijet and $Z+$jets samples are both shown. The first uncertainty is statistical while the second is systematic.}
%\label{fake_result}
%\end{table}
%\renewcommand{\arraystretch}{1.0}

\section{Systematic uncertainties}  \label{syst}

Relative uncertainties on the expected yields of the dominant $WZ$ background 
and the EGM $W'$ signal with $m_{W'}=1$ \TeV~in \srhm~are shown in
Table~\ref{uncertainties}. These uncertainties are representative of those found for other signal masses and background types.
The lepton-related ones include
%standard set of systematic uncertainties relating to object
%reconstruction are computed, based on performance studies of the ATLAS detector. 
uncertainties from the lepton trigger, identification, energy scale, energy resolution, isolation, and impact parameters. 
% uncertainties, which, for the SM $WZ$ process
%(similar values are obtained for the other backgrounds), tally to a combined uncertainty of 3.1\%,
%2.1\%, 1.8\%, and 1.9\% in the $e\nu ee$, $\mu\nu ee$, $e\nu \mu\mu$, and
%$\mu\nu\mu\mu$ channels respectively. Moreover, given that \etmiss~is computed using 
%reconstructed objects, each of their uncertainties including jet energy scale and resolution are propagated to 
%the \etmiss~calculation. These latter two uncertainties amount to an additional uncertainty of 
%2.0\% on the inclusive yields in \srhm, again for the SM $WZ$
%production. However, the dominant uncertainty on the measurement of \etmiss~comes from the
%effects of pileup, which translate to an average uncertainty of
%1.5\% in absolute yields within the four channels in \srhm. 
The uncertainties on the lepton momentum and jet energy scales and resolutions are propagated to the \etmiss calculation.
Other \etmiss-related uncertainties include those on soft energy deposits due to additional $pp$ collisions, and energy deposits 
%around clusters associated to reconstructed jets and electrons, energy deposits not associated to any reconstructed 
%associated to reconstructed jets and energy deposits 
not associated with any reconstructed object.
Both the normalization and shape uncertainties are taken into account from the above sources.
%The uncertainties on both the normalization and shape from the above sources are taken into consideration 
%for all MC estimations of the backgrounds and signals. 
%, and lepton momentum measurement. 

Cross-section uncertainties for the dominant SM physics processes are computed via
\texttt{MCFM}~\cite{mcfm_VV}, which provides NLO QCD calculations for diboson production cross sections. 
The relative uncertainty due to higher-order corrections to the $WZ$ cross 
sections is 5\%~\cite{nNLO}. 
The renormalization and factorization scales are varied by a factor of
two relative to their nominal values. The resulting sum in quadrature of
the uncertainties in \srhm~on the $WZ$, $ZZ$, and $Z\gamma$ cross sections are found
to be 6.9\%, 4.3\%, and 5.0\% respectively. 
%The renormalization and factorization scale uncertainties are calculated independently and added in quadrature. 
PDF uncertainties are derived by comparing the predicted cross
sections using the NLO CT10 and 
MSTW PDF as well as the CT10 eigenvector error PDF sets (90\% confidence level). The resulting uncertainties 
are 4.1\%, 4.7\% and 3.2\% for these three processes respectively. 
%of, on average, 3\% are also included, while statistical uncertainties
%on these MC predictions in \srhm are 2.2\% of the total event yields. 

%samples have been generated with $m_{W'}$ from 200
%\GeV to 2 \TeV and the acceptance times efficiencies for these signals are
%shown in figure~\ref{wprime_acc}. The statistical uncertainties vary
%from 8\% to 2\% as a function of $m_{WZ}$.
%All systematic uncertainties, including lepton ID (2-3\%),
%reconstruction efficiencies (1-2\%), jet energy scale and resolution
%(2\%) and PDF (3\%) uncertainties are individually interpolated,
%thereby interpolating signal shape uncertainties as well.    

%Given that the SM background modelling suffers from low MC event counts  
%in the tail of the $m_{WZ}$ distribution, an extrapolation method is
%devised to smooth the predicted yields. The method consists in
%fitting independently the $WZ$ background in the region 
%with $m_{WZ}>500$ \GeV and the non-$WZ$ backgrounds in the region with $m_{WZ}>300$ \GeV, 
%each with a power law function. The uncertainty on these two fits dominates all other uncertainties in the
%$m_{WZ}$ range beyond 800 \GeV (50\% at 800 \GeV to 200\% at 2 \TeV). 

Given that the SM background modelling suffers from low MC event counts
in the tail of the $m_{WZ}$ distribution, an extrapolation method is
devised to smooth the predicted yields. The method consists in
performing two independent $\chi^2$ fits, one on the $WZ$ background in the region
with $m_{WZ}>500$ \GeV, and a second on the sum of all non-$WZ$ backgrounds in the region with $m_{WZ}>300$ \GeV,
each with the power-law function $N(x) = c_0 x^{c_1}$, 
%where $x$ is the $m_{WZ}$ invariant mass. 
where $x$ is $m_{WZ}$.
%The normalization parameter $c_0$ is adjusted and
The overall normalization of the fitted function is 
set to the expected number of events for each of the two types of background. 
The non-$WZ$ backgrounds are fitted jointly to gain from
their combined size, thus reducing the total uncertainty
in the fit, which is computed via the minimization function's Hessian error matrix.
Other fitting functions such as
an exponential or more elaborate power-law functions were tested,
but their shapes were found to be within the uncertainties from the simple power-law
function given above. Hence, only the uncertainties from the simple power-law function are considered, 
and these dominate all other uncertainties in the
range $m_{WZ} > 800$~\GeV~(e.g. the fit uncertainty reaches 50\%
of the total expected yields at $m_{WZ} = 800$~\GeV, and 400\% at $m_{WZ}
= 1.6$~\TeV). 

Additionally, the shapes of the $m_{WZ}$ distribution for the SM $WZ$ process predicted by \powheg~and the
multi-leg generators \sherpa~and \madgraph, as well as NLO generators such as \mcatnlo~are compared. The largest
deviations from the \powheg~distribution are used as systematic uncertainties on the predicted $m_{WZ}$ shape.  

 \renewcommand{\arraystretch}{1.2}
 \begin{table*}[htbp]
 \caption{Relative uncertainties in the expected yields for the SM $WZ$ background and the EGM $W'$ signal with $m_{W'}=1$ \TeV~in the high-mass signal region (SR$_\textrm{HM}$).
The renormalization and factorization scales, together with the PDF uncertainties on the fiducial cross section are included under theoretical uncertainty for SM $WZ$ background.
For EGM $W'$ signal, the theoretical uncertainty stands for the effects of the scale
and PDF uncertainties, added in quadrature, on its acceptance.
Shape-related uncertainties are not included here. Similar results are found in the low-mass signal region (SR$_\textrm{LM}$).} \vspace*{1mm}
\label{uncertainties}
 \centering
% \footnotesize
 \begin{tabular}{c|cccc|cccc} 
 \hline
 \hline
  \multicolumn{1}{c|} {Uncertainty} & \multicolumn{4}{c}{SM $WZ$} & \multicolumn{4}{|c}{EGM $W'$ ($m_{W'}=1$ \TeV)} \\
 \cline{2-9}
  sources & $e\nu ee$ & $\mu\nu ee$ & $e\nu\mu\mu$ & $\mu\nu\mu\mu$ & $e\nu ee$ & $\mu\nu ee$ & $e\nu\mu\mu$ & $\mu\nu\mu\mu$ \\
 \hline
 %MC statistical & 2.7\% & 2.0\% & 2.0\% & 2.2\% & 2.5\% & 2.5\% & 2.5\% & 2.5\% \\
 MC statistics & 2.7\% & 2.0\% & 2.0\% & 2.2\% & 2.5\% & 2.5\% & 2.5\% & 2.5\% \\
 Lepton-related & 3.1\% & 1.8\% & 1.8\% & 1.9\% & 3.7\% & 2.6\% & 2.1\% & 2.4\% \\
 \etmiss-related & 2.8\% & 1.9\% & 2.6\% & 1.7\% & 1.1\% & 0.4\% & 0.4\% & 0.4\% \\
 Luminosity & 2.8\% & 2.8\% & 2.8\% & 2.8\% & 2.8\% & 2.8\% & 2.8\% & 2.8\% \\
 %Theoretical & 9.5\% & 9.5\% & 9.5\% & 9.5\% & 0.6\% & 0.5\% & 0.2\% & 0.2\% \\
 Theory & 9.5\% & 9.5\% & 9.5\% & 9.5\% & 0.6\% & 0.5\% & 0.2\% & 0.2\% \\
% Total & .7 & 2.0 & 2.0 & 2.2 & 2.7 & 2.0 & 2.0 & 2.2 \\
 \hline
 \hline
 \end{tabular}
 \end{table*}
% \renewcommand{\arraystretch}{1.2}

%The EGM $W'$ signal samples are generated with $m_{W'}$ from 200 \GeV to 
%2 \TeV with a step size of 200 \GeV. %The combined acceptance times efficiency ($A \times \epsilon$) as a function of 
%the $W'$ mass is shown in Fig.~\ref{wprime_acc} for the two signal regions separately. 
%The acceptance times efficiency ($A \times \epsilon$) in \srhm increases from 2\% to 30\% for $m_{W'}$ from 200 \GeV to 2 \TeV, 
%while the $A \times \epsilon$ in \srlm decreases from 8\% to 0\% in this region. 
A procedure was developed to obtain the $m_{WZ}$ distribution for any given $m_{W'}$ mass point using a functional interpolation between the available $m_{WZ}$ signal templates. These distributions are individually fitted with a crystal ball function using \roofit~\cite{RooFIT}. The 4 crystal ball parameters are then each fitted as a function of the $W^{\prime}$ mass to build the $m_{WZ}$ template for any intermediate $W^{\prime}$ mass point. All systematic uncertainties are individually interpolated. 
%thereby interpolating signal shape uncertainties as well.    

%\begin{figure}[htp]
%\includegraphics[width=\columnwidth]{figures/acc_vs_mass_SR_WZCR_final.eps}
%\caption{Acceptance times efficiency of the EGM $W^{\prime}$ boson in the two signal regions for four channels combined. 
 % Error bars reflect the statistical uncertainties of each samples.} \label{wprime_acc}
%\end{figure}

Theoretical uncertainties on the EGM $W'$ signal yields primarily come from uncertainties 
on the reconstructed signal's acceptance times efficiency %($A \times \epsilon$)
due to the PDF set used. 
%Similar to the PDF uncertainty evaluation  for the SM physics processes, 
%The PDF uncertainties on the signal acceptance are
The uncertainties in the signal acceptance due to the PDF are
derived from the MSTW eigenvector error sets, and the difference
between the predictions of the CT10 and MSTW PDF sets, combined in quadrature.

\section{Results} \label{results}

\begin{figure}[hbtp]
\centering
\includegraphics[width=\columnwidth]{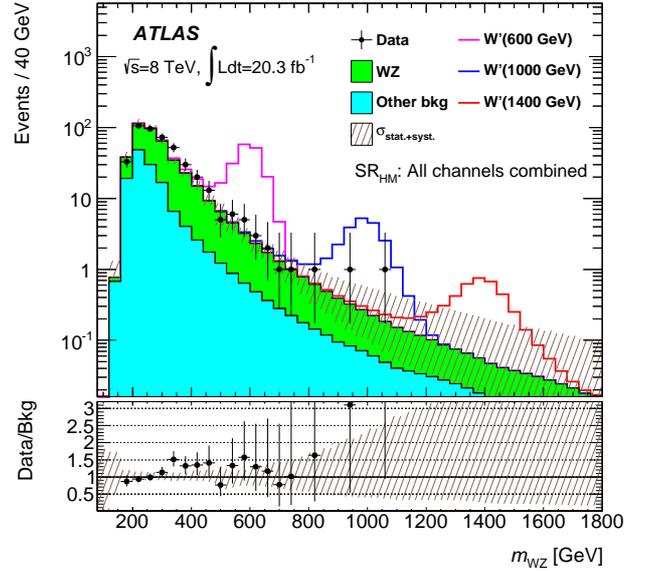}
\caption{Observed and predicted $WZ$ invariant mass ($m_{WZ}$) distribution for events in the high-mass signal region (SR$_\textrm{HM}$). 
 An extrapolation of the backgrounds to the very-high-mass region was performed 
 using a power-law function to fit for the SM $WZ$ and the sum of all other backgrounds separately. 
 Predictions from $W'$ samples with masses of 600~\GeV, 1000~\GeV~and 1400~\GeV~are also
 shown, stacked on top of the expected backgrounds. 
%Statistical and
% systematic uncertainties are respectively shown throughout.
The uncertainty bands upon the expected background include
both the statistical and systematic uncertainties in the MC simulation and the fake-background estimation added in quadrature.
} \label{mWZ_log}
\end{figure}

\begin{figure}[hbtp]
\centering
\includegraphics[width=\columnwidth]{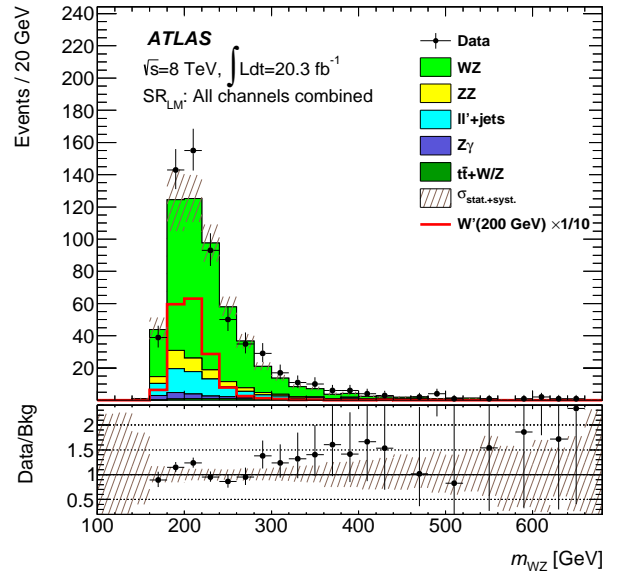}
\caption{Observed and predicted $WZ$ invariant mass ($m_{WZ}$) distribution for events in the
  low-mass signal region (SR$_\textrm{LM}$). Predictions from a $W'$ sample with mass of 200 \GeV~are also shown.
  The $W'$ curve is scaled by $1/10$ for better display. 
The uncertainty bands upon the expected background include
both the statistical and systematic uncertainties in the MC simulation and the fake-background estimation added in quadrature.
} \label{mWZ_log_lmsr}
  %\srlm. } \label{mWZ_log_lmsr}
\end{figure}

The $m_{WZ}$ spectrum in the two signal regions is scrutinized for
excesses of data over the predicted SM backgrounds. A total of 449 $WZ$ candidate events in \srhm~are observed in the data after applying all event selection criteria, to be compared with the SM prediction of 
%$420.9 \pm 4.5 {\textrm{(stat.)}} ^{+56.3}_{-39.0} {\textrm{(syst.)}}$ events. The corresponding numbers in \srlm are 617 events selected in data and 
%$562.7 \pm 5.1 {\textrm{(stat.)}} ^{+54.7}_{-42.6} {\textrm{(syst.)}}$ events expected from SM processes. 
$421 \pm 5 {\textrm{(stat.)}} ^{+56}_{-39} {\textrm{(syst.)}}$ events. The corresponding numbers in \srlm~are 617 events selected in the data and 
$563 \pm 5 {\textrm{(stat.)}} ^{+55}_{-43} {\textrm{(syst.)}}$ events expected from SM processes. 
The observed $m_{WZ}$ distribution in \srhm~is compared to the expected SM background distribution in Fig.~\ref{mWZ_log}, which 
combines all four lepton decay channels. The contributions from hypothetical 
EGM $W'$ bosons with masses of 600, 1000, and 1400~\GeV~are also shown. %and scaled to LO in cross section. 
A breakdown of the signal, backgrounds, and observed data yields in \srhm~is shown in Table~\ref{final_yields} for 
each individual channel and also for all four channels combined.
The $m_{WZ}$ distribution in \srlm~is shown in
Fig.~\ref{mWZ_log_lmsr}. 

\begin{table*}[htbp]
  \caption{The estimated background yields, the observed number of data events, and the predicted signal yield for a set of $W'$ resonance masses in
the high-mass signal region (SR$_\textrm{HM}$).} \vspace*{1mm}
    \label{final_yields}
\centering
\resizebox{\textwidth}{!}{
\renewcommand{\arraystretch}{1.2}
\begin{tabular}{l|c|c|c|c|c}
\hline
\hline
            &  $e\nu ee$       &  $\mu\nu ee$  &  $e\nu\mu\mu$  &   $\mu\nu\mu\mu$   & Combined \\
\hline
Backgrounds:   &  &  &  &  & \\
%~$WZ$                           & $56.5 \pm 1.5 \pm 6.1 $  & $68.6 \pm 1.4 \pm 7.0$  & $70.1 \pm 1.4 \pm 7.2$  & $89.8 \pm 2.0 \pm 9.1 $  & $285.0 \pm 3.2 \pm 28.8$ \\
~$WZ$                           & $56.5 \pm 1.5 \pm 6.1 $  & $68.6 \pm 1.4 \pm 7.0$  & $70.1 \pm 1.4 \pm 7.2$  & $89.8 \pm 2.0 \pm 9.1 $  & $285 \pm 3 \pm 29$ \\
~$ZZ$                           & $8.7 \pm 0.1 \pm 0.9  $  & $8.7 \pm 0.2 \pm 0.8 $  & $11.7 \pm 0.2 \pm 1.3$  & $11.6 \pm 0.2 \pm 1.1$   & $40.7 \pm 0.4 \pm 3.9$ \\
~$Z\gamma$                      & $6.4 \pm 0.8 \pm 1.5 $   & $<0.05$                 & $8.1 \pm 0.9 \pm 1.2 $  & $<0.05$                  & $14.5 \pm 1.2 \pm 2.2 $ \\
~$t\bar{t}+W/Z$			        & $2.5 \pm 0.1 \pm 0.8 $   & $3.2 \pm 0.1 \pm 1.0 $  & $2.6 \pm 0.1 \pm 0.8 $  & $3.3 \pm 0.1 \pm 1.0$    & $11.6 \pm 0.2 \pm 3.5$ \\

%~Fake Background  & $12.7 \pm 1.0^{+8.9}_{-5.6}$  & $19.3 \pm 1.7^{+10.6}_{-4.1}$  & $14.0 \pm 1.0^{+12.6}_{-6.9}$  & $23.1 \pm 1.9^{+14.9}_{-7.2}$  & $69.1 \pm 2.9^{+47.0}_{-23.8}$\\ \hline
~$\ell\ell'+$jets & $12.7 \pm 1.0^{+8.9}_{-5.6}$  & $19 \pm 2^{+11}_{-4}$  & $14 \pm 1^{+13}_{-7}$  & $23 \pm 2^{+15}_{-7}$  & $69 \pm 3^{+47}_{-24}$\\
 \hline
%Sum of Backgrounds               & $86.8 \pm 2.0^{+11.3}_{-8.9}$ & $99.8 \pm 2.2^{+12.9}_{-8.4}$  & $106.5 \pm 2.0^{+15.0}_{-10.7}$  & $127.8 \pm 2.8^{+17.6}_{-11.9}$  & $420.9 \pm 4.5^{+56.3}_{-39.0}$ \\
Sum of Backgrounds               & $87 \pm 2^{+11}_{-9}$ & $100 \pm 2^{+13}_{-8}$  & $107 \pm 2^{+15}_{-11}$  & $128 \pm 3^{+18}_{-12}$  & $421 \pm 5^{+56}_{-39}$ \\
Data                             & $99$  & $90$  & $136$ & $124$  & $449$ \\ \hline
Signals: & & & & \\
%@LO
%~~$W^{'} \rightarrow WZ$ ($M(W^{'})=600$~\GeV)  & $42.0 \pm 1.2 \pm 2.1$ & $48.3 \pm 1.3 \pm 2.4$ & $46.5 \pm 1.3 \pm 2.3$ & $52.9 \pm 1.4 \pm 2.6$ & $189.7 \pm 2.6 \pm 9.5$ \\
%~~$W^{'} \rightarrow WZ$ ($M(W^{'})=1000$~\GeV) & $5.7 \pm 0.1 \pm 0.3$ & $5.9 \pm 0.2 \pm 0.3$ & $5.7 \pm 0.1 \pm 0.3$ & $5.7 \pm 0.1 \pm 0.3$ & $23.0 \pm 0.3 \pm 1.0$ \\
%~~$W^{'} \rightarrow WZ$ ($M(W^{'})=1400$~\GeV) & $1.1 \pm 0.1 \pm 0.1$ & $1.1 \pm 0.1 \pm 0.1$ & $1.0 \pm 0.1 \pm 0.1$ & $1.0 \pm 0.1 \pm 0.1$ & $4.2 \pm 0.2 \pm 0.2$ \\ 
%@NNLO
~~$W^{'} \rightarrow WZ$ ($M(W^{'})=600$~\GeV)  & $54.2 \pm 1.6 \pm 2.7$ & $62.2 \pm 1.7 \pm 3.1$ & $59.9 \pm 1.7 \pm 3.0$ & $68.2 \pm 1.8 \pm 3.4$ & $244 \pm 3 \pm 12$ \\
~~$W^{'} \rightarrow WZ$ ($M(W^{'})=1000$~\GeV) & $7.1 \pm 0.2 \pm  0.4$ & $7.4 \pm 0.2 \pm 0.4$  & $7.1 \pm 0.2 \pm 0.4$ & $7.1 \pm 0.2 \pm 0.4$ & $28.6 \pm 0.4 \pm 1.3$ \\
~~$W^{'} \rightarrow WZ$ ($M(W^{'})=1400$~\GeV) & $1.3 \pm 0.1 \pm 0.1$ & $1.3 \pm 0.1 \pm 0.1$ & $1.3 \pm 0.1 \pm  0.1$ & $1.2 \pm 0.1 \pm 0.1$ & $5.1 \pm 0.1 \pm 0.2$ \\ 
%
%~~$W^{'} \rightarrow WZ$ ($M(W^{'})=600$~\GeV)  & $54.15 \pm 1.55 \pm 2.7$ & $62.21 \pm 1.68 \pm 3.1$ & $59.87 \pm 1.66 \pm 3.0$ & $68.17 \pm 1.78 \pm 3.4$ & $244.4 \pm 3.3 \pm 12.2$ \\
%~~$W^{'} \rightarrow WZ$ ($M(W^{'})=1000$~\GeV) & $7.09 \pm 0.18 \pm  0.4$ & $7.35 \pm 0.18 \pm 0.4$  & $7.06 \pm 0.18 \pm 0.4$ & $7.06 \pm 0.18 \pm 0.4$ & $28.6 \pm 0.36 \pm 1.3$ \\
%~~$W^{'} \rightarrow WZ$ ($M(W^{'})=1400$~\GeV) & $1.29 \pm 0.032 \pm 0.1$ & $1.34 \pm 0.032 \pm 0.1$ & $1.27 \pm 0.03 \pm  0.1$ & $1.19 \pm 0.03 \pm 0.1$ & $5.09 \pm 0.1 \pm 0.2$ \\ 

\hline
\hline
  \end{tabular}}
\end{table*}

The $m_{WZ}$ distribution is used to build a binned log-likelihood ratio (LLR) test statistic~\cite{llr1967}.
%The background is fixed as estimated. 
The systematic uncertainties are represented by nuisance parameters for both the backgrounds and signals.
Confidence levels (CL) for the signal-plus-background hypothesis ($\textrm{CL}_{\textrm{s+b}}$) and background-only hypothesis ($\textrm{CL}_{\textrm{b}}$) 
are computed by integrating the LLR distributions obtained from simulated pseudo-experiments using Poisson 
statistics. 

To check the consistency between the observed data and expected SM
backgrounds, the $p$-value, defined as
$1-\textrm{CL}_{\textrm{b}}$, for a background 
fluctuation to give rise to an excess at least as large as that observed
in data is computed. The obtained $p$-values are reported in Table~\ref{limit_table} for the
signal hypothesis of a $W^{\prime}$ particle with mass from 200 \GeV~to
2 \TeV. The lowest local $p$-value probability is found to be 8\% 
%in the combined channel 
for the 375~\GeV~resonance mass hypothesis, equivalent to a 1.75$\sigma$ local excess, % {\it \bf (to be indicating that the observed data are consistent with the SM background
%expectations, 
indicating that no significant excess is observed. 

In the modified frequentist approach~\cite{CLs}, the 95\% CL excluded cross section is computed as the cross section for which 
 $\textrm{CL}_{\textrm{s}}$, defined as the ratio $\textrm{CL}_{\textrm{s+b}}/\textrm{CL}_{\textrm{b}}$, 
is equal to 0.05. 
%The limits statistically combine all lepton
%decay channels 
%as well as the two signal regions to maximize the
%sensitivity of the search. 
For the mass points above 400 \GeV, only the high-mass signal region is used in the calculation by statistically combining all lepton decay channels.
For the mass points below or equal to 400 \GeV, the two signal regions are further combined to maximize
the sensitivity of the search. 

%To check the consistency between the observed data and expected SM backgrounds, the probability that a background 
%fluctuation gives rise to an excess at least as large as that observed in data is computed and reported in Table~\ref{limit_table} for the signal hypothesis of a $W^{\prime}$
%particle with mass from 200 \GeV to 1 \TeV. The lowest local $p$-value probability is found to be 8.0\% in the combined channel for the 375~\GeV resonance mass hypothesis, equivalent to a 1.75$\sigma$ local excess, % {\it \bf (to be 
% indicating that the observed data are consistent with the SM background
%expectations, and that therefore no significant excess is observed.

\begin{figure}[hbtp]
\centering
\includegraphics[width=\columnwidth]{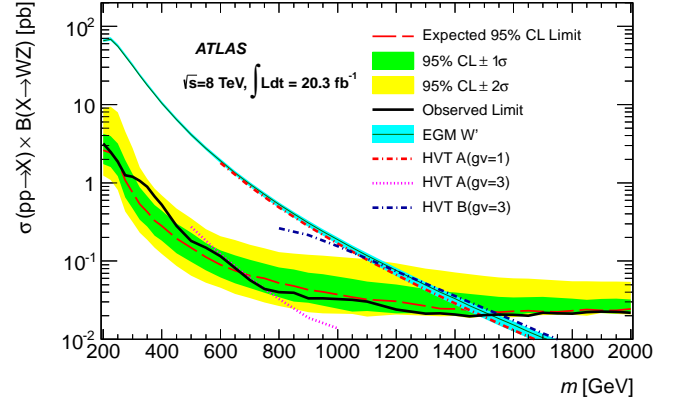}
%\includegraphics[width=\columnwidth]{figures/Preliminary/limitplot_wprimeincl.eps}
%\caption{The observed 95\% CL upper limits on $\sigma(pp \rightarrow W') \times \textrm{BR}(W' \rightarrow WZ)$ as a function of 
\caption{
%The observed 95\% CL upper limits on $\sigma(pp \rightarrow W') \times \textrm{B}(W' \rightarrow WZ)$ as a function of 
The observed 95\% CL upper limits on $\sigma(pp \rightarrow X) \times \textrm{B}(X \rightarrow WZ)$ as a function of 
the signal mass $m$, where $X$ stands for the signal resonance. 
The expected limits are also shown together with the
$\pm 1$ and $\pm 2$ standard deviation uncertainty bands. 
Both the expected and observed upper limits assume the EGM $W'$ signal acceptance
times efficiency as presented in Table~\ref{limit_table}. 
Theoretical cross sections for the EGM $W'$ and the HVT benchmark models are also
shown. The uncertainty band around the EGM $W^{\prime}$ cross-section
line represents the theoretical uncertainty on the NNLO cross-section calculation using ZWPROD~\cite{ZWPROD}.} \label{limit}
\end{figure}

\begin{figure}[hbtp]
\centering
\includegraphics[width=\columnwidth]{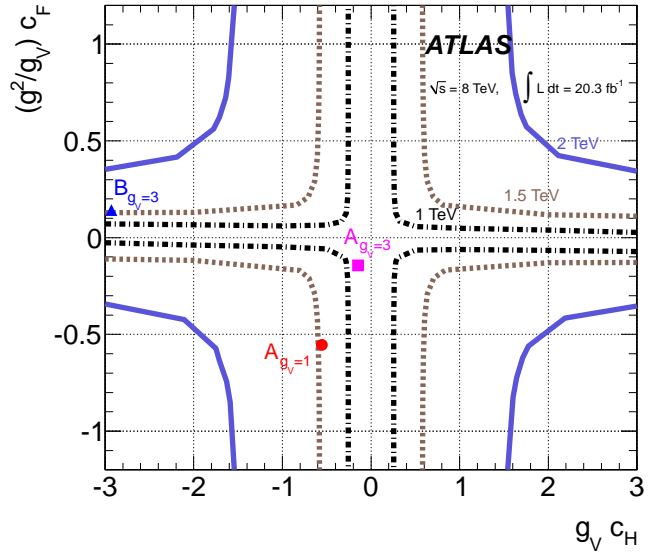}
\caption{Observed 95\% CL exclusion contours in the HVT parameter space $\{(g^2/g_V)c_F, g_Vc_H\}$ for resonances of mass 1~\TeV,
1.5~\TeV and 2~\TeV. Also shown are the benchmark model parameters $A_{(g_V=1)}$ (circle) and $A_{(g_V=3)}$ (square) and $B_{(g_V=3)}$ (triangle).} \label{HVT_exclusion}
\end{figure}

%Figure~\ref{limit} presents the 95\% CL upper limits on $\sigma(pp \rightarrow W') \times \textrm{BR}(W' \rightarrow WZ)$ 
%Fig.~\ref{limit} presents the 95\% CL upper limits on $\sigma(pp \rightarrow W') \times \textrm{B}(W' \rightarrow WZ)$ 
%as a function of the EGM $W'$ mass together with the theoretical cross sections of the EGM $W'$ and HVT benchmark models. 
Fig.~\ref{limit} presents the 95\% CL upper limits on $\sigma(pp \rightarrow X) \times \textrm{B}(X \rightarrow WZ)$ 
as a function of the signal resonance mass, where $X$ stands for the signal resonance, together with the theoretical cross sections of the EGM $W'$ and HVT benchmark models. 
The latter cross sections are calculated via the web interface~\cite{HVT-web} provided by the authors of Ref.~\cite{Pappadopulo:2014qza}.
The exclusion region in parameter space $\{(g^2/g_V)c_F, g_V c_H\}$ is shown in Fig.~\ref{HVT_exclusion}. 
The fermion coupling $c_F$ was set to the same value for quarks and leptons. The couplings
$c_{VVV}$, $c_{VVHH}$ and $c_{VVW}$, which involve vertices with more than one heavy vector boson and which have negligible effect on the cross section, were set to zero. 
Table~\ref{limit_table} presents the expected and observed limits for a selected set of signal mass points 
as well as the EGM $W'$ signal acceptance $A$ and correction
factor $C$. 
The acceptance $A$ is defined as the number of
generated events found within the fiducial region at particle level divided by the total
number of generated events, while 
$C$ is defined as the number of reconstructed events passing the
nominal selection requirements divided by the number of generated
events within the fiducial region at particle level. The fiducial region selection criteria consist of the same kinematic selections (lepton $p_{\rm T}$, lepton $\eta$, $Z$ boson mass, $E^{\textrm{miss}}_{\rm T}$, $\Delta y(W,Z)$ and $\Delta \phi(\ell, E^{\textrm{miss}}_{\rm T})$) and lepton isolation requirements as in the nominal selections.
Particle level refers to particle states that stem from the hard scatter, including those that are
the product of hadronization, but before their interaction with the detector. 
%either the direct product of the hard scatter or the product of hadronization. 
% where the latter is defined as the number of reconstructed events passing the nominal
%selection over the number of events passing the same selection at parton level. 
%Note that $A \times C$ = $\mathcal{A} 
%\times \epsilon$, where $\mathcal{A}$ and $\epsilon$ are respectively the reconstructed 
%acceptance and efficiency.
Table~\ref{t-limits} presents the 95\% CL expected and observed lower
limits on the EGM $W^{\prime}$~boson mass for each decay channel and
their combination. 
The observed
(expected) exclusion limit on the EGM $W^{\prime}$ mass is found to be 
%1525 (1511)~\GeV. 
1.52 (1.49)~\TeV, and the limits in each channel are shown in Table~\ref{t-limits}.
%, which improves on the current published limits by over 300~\GeV.
%boson with masses below  The observed band reaches $+2.x \sigma$ for the 350~\GeV signal mass
%hypothesis. The theoretical curves' intersections with the observed
%(expected) limit produce mass limits of 1500 (1400)~\GeV for the
%EGM $W'$ and 
The simulated HVT resonances are found to have kinematic distributions
similar to those of the $W^{\prime}$ and thus have similar acceptances to the EGM model. The corresponding observed (expected) limits for
the $A(g_V = 1)$, $A(g_V = 3)$, and $B(g_V = 3)$ HVT
resonances from Ref.~\cite{Pappadopulo:2014qza} 
%are 1491 (1453)~\GeV, 763 (689)~\GeV, and 1557 (1532)~\GeV respectively. 
are 1.49 (1.45)~\TeV, 0.76 (0.69)~\TeV, and 1.56 (1.53)~\TeV~
respectively. In Fig.~\ref{limit}, the HVT benchmark model curves are not shown for low
resonance mass where the models do not apply. 

\begin{table*}[htbp]  
    \caption{The expected and observed 95\% CL upper limits on the production cross section of narrow resonances decaying
      to $WZ$ as a function of their mass. The high-mass signal region (SR$_\textrm{HM}$)  and low-mass signal region (SR$_\textrm{LM}$)  fiducial acceptances at
      particle level ($A$) and correction factors ($C$) for a EGM $W'$ as
      implemented in \pythia~are also given.
      %The mass points beyond 400~\GeV in \srlm were not used in the limit setting due to their very low acceptances. 
      \srlm~was not used in setting the limits for the mass points beyond 400~\GeV~due to their very low acceptances.
      Errors shown are statistical.
      The $p$-value, defined as $1-\textrm{CL}_{\textrm{b}}$, is also shown for each mass point in the last column.}\vspace*{1mm}
    \label{limit_table}
%  \resizebox{\textwidth}{!}{   
    \centering
\renewcommand{\arraystretch}{1.2}
    \begin{tabular}{c|cc|c|c|c}
      \hline
      \hline
      $m_{W'}$ & \multicolumn{2}{c|}{Excluded $\sigma \times \textrm{B}$ [fb]} & \srhm & \srlm  & \multirow{2}{*}{$p$-value}\\
      $[${\GeV}$]$    & Expected & Observed   & $A/C$ & $A/C$  \\
      \hline
      200 & 2613 & 3182  & $0.025 \pm 0.001$ / $0.75 \pm 0.05$ &  $0.135 \pm 0.003$ / $0.57 \pm 0.02$  & 0.36 \\
      250 & 1902 & 1853  & $0.111 \pm 0.002$ / $0.55 \pm 0.02$ & $0.070 \pm 0.002$ / $0.80 \pm 0.03$ & 0.48 \\
      300 & 751  & 1195  & $0.202 \pm 0.003$ / $0.57 \pm 0.01$ & $0.024 \pm 0.001$ / $1.42 \pm 0.07$ & 0.22 \\
      350 & 427  & 894   & $0.269 \pm 0.004$ / $0.61 \pm 0.01$ & $0.0093 \pm 0.0006$ / $2.5 \pm 0.2$ & 0.094 \\
      375 & 330  & 670   & $0.29 \pm 0.01$   / $0.62 \pm 0.02$   & $0.007 \pm 0.001$ / $ 2.9 \pm 0.6$ & 0.080 \\
      400 &  281 &  526 & $0.311 \pm 0.005$ / $0.63 \pm 0.01$ & $0.0048 \pm 0.0005$ / $3.3 \pm 0.4$ & 0.094 \\
      600 &  90 &  115 & $0.426 \pm 0.006$ / $0.68 \pm 0.01$ &      \multirow{8}{*}{{\it not used}} & 0.29 \\
      800 &  52 &  40 & $0.475 \pm 0.006$ / $0.68 \pm 0.01$ &  &  0.71\\
      1000 &  38 & 33 & $0.505 \pm 0.007$ / $0.68 \pm 0.01$ &   & 0.59 \\
      1200 &  31 & 24 & $0.526 \pm 0.007$ / $0.66 \pm 0.01$ &  & 0.71 \\
      1400 &  25 & 21 & $0.530 \pm 0.007$ / $0.66 \pm 0.01$ & & 0.81 \\
      1600 &  23 & 21 & $0.533  \pm 0.007$ / $0.63 \pm 0.01$ & &  0.83\\
      1800 &  23 & 21 & $0.544 \pm 0.007$ / $0.60 \pm 0.01$ & &  0.82\\
      2000 &  24 & 22 & $0.535 \pm 0.007$ / $0.57 \pm 0.01$ &  &  0.85\\
      \hline
      \hline
    \end{tabular}%}
  \end{table*}

%\FloatBarrier

\begin{table}[htbp]
\caption{Expected and observed lower mass limits at 95\% CL in \TeV~for the EGM $W'$~boson in
  the $e\nu ee$, $e\nu\mu\mu$, $\mu\nu ee$, $\mu\nu\mu\mu$ channels as well as the four channels combined.}\vspace*{1mm}
  \label{t-limits}
\begin{center}{%\small
\renewcommand{\arraystretch}{1.2}
  \begin{tabular}{c|cccc|c}
\hline
\hline
    & \multicolumn{5}{c}{Excluded EGM $W'$ lower mass $[${\TeV}$]$} \\
 \cline{2-6}
    & $e\nu ee$ & $\mu\nu ee$  & $e\nu\mu\mu$ & $\mu\nu\mu\mu$ & combined\\
\hline
%Expected     & 1208     & 1162       & 1173         & 1155     & 1489  \\
%Observed     & 1204     & 1188       & 1061          & 1170     & 1522   \\
Expected & 1.21     & 1.16       & 1.17         & 1.16     & 1.49  \\
Observed & 1.20     & 1.19       & 1.06         & 1.17     & 1.52   \\
\hline
\hline
  \end{tabular}}
\end{center}
\end{table}

\section{Conclusion}
A search for resonant $WZ$~diboson production in the fully leptonic
channel has been performed with the ATLAS
detector, using \lumi~fb$^{-1}$ of $pp$ collision data collected at
$\sqrt{s}=8$ \TeV~at the LHC. 
No excess is found in data compared to the SM expectations. 
Stringent limits on the production cross section times $WZ$ branching
ratio are 
%obtained for a $W^{\prime}$ arising from an extended gauge model
%decaying to $WZ$ as a function of the
%resonance mass. 
obtained as a function of the resonance mass for a $W^{\prime}$ arising from an extended gauge model and decaying to WZ.
A corresponding observed (expected) mass limit of 1.52 (1.49) ~\TeV~is derived for the $W^{\prime}$.
%The EGM $W'$ boson is used as a benchmark model and 
%stringent limits are set 
%on the $W'$ and HVT $V$ production cross section times branching ratio to $WZ$ as a
%function of the resonance mass.

%\documentclass[11pt,a4paper,dvips]{article}

%\begin{document}

% Acknowledgements for papers with collision data
% Version 19-Feb-2014

\section*{Acknowledgements}

% Standard acknowledgements start here
%----------------------------------------------
We thank CERN for the very successful operation of the LHC, as well as the
support staff from our institutions without whom ATLAS could not be
operated efficiently.

We acknowledge the support of ANPCyT, Argentina; YerPhI, Armenia; ARC,
Australia; BMWF and FWF, Austria; ANAS, Azerbaijan; SSTC, Belarus; CNPq and FAPESP,
Brazil; NSERC, NRC and CFI, Canada; CERN; CONICYT, Chile; CAS, MOST and NSFC,
China; COLCIENCIAS, Colombia; MSMT CR, MPO CR and VSC CR, Czech Republic;
DNRF, DNSRC and Lundbeck Foundation, Denmark; EPLANET, ERC and NSRF, European Union;
IN2P3-CNRS, CEA-DSM/IRFU, France; GNSF, Georgia; BMBF, DFG, HGF, MPG and AvH
Foundation, Germany; GSRT and NSRF, Greece; ISF, MINERVA, GIF, I-CORE and Benoziyo Center,
Israel; INFN, Italy; MEXT and JSPS, Japan; CNRST, Morocco; FOM and NWO,
Netherlands; BRF and RCN, Norway; MNiSW and NCN, Poland; GRICES and FCT, Portugal; MNE/IFA, Romania; MES of Russia and ROSATOM, Russian Federation; JINR; MSTD,
Serbia; MSSR, Slovakia; ARRS and MIZ\v{S}, Slovenia; DST/NRF, South Africa;
MINECO, Spain; SRC and Wallenberg Foundation, Sweden; SER, SNSF and Cantons of
Bern and Geneva, Switzerland; NSC, Taiwan; TAEK, Turkey; STFC, the Royal
Society and Leverhulme Trust, United Kingdom; DOE and NSF, United States of
America.

The crucial computing support from all WLCG partners is acknowledged
gratefully, in particular from CERN and the ATLAS Tier-1 facilities at
TRIUMF (Canada), NDGF (Denmark, Norway, Sweden), CC-IN2P3 (France),
KIT/GridKA (Germany), INFN-CNAF (Italy), NL-T1 (Netherlands), PIC (Spain),
ASGC (Taiwan), RAL (UK) and BNL (USA) and in the Tier-2 facilities
worldwide.
%----------------------------------------------
%\end{document}

\section*{References}
\bibliographystyle{elsarticle-num}
\bibliography{Paperlvll}

\onecolumn
\clearpage
% ATLAS Collaboration author list
% Data extracted on 12-May-2014 for paper reference EXOT-2013-07
%\documentclass[11pt]{article}
%\usepackage{a4wide}\begin{document}
\begin{flushleft}
{\Large The ATLAS Collaboration}

\bigskip

G.~Aad$^{\rm 84}$,
B.~Abbott$^{\rm 112}$,
J.~Abdallah$^{\rm 152}$,
S.~Abdel~Khalek$^{\rm 116}$,
O.~Abdinov$^{\rm 11}$,
R.~Aben$^{\rm 106}$,
B.~Abi$^{\rm 113}$,
M.~Abolins$^{\rm 89}$,
O.S.~AbouZeid$^{\rm 159}$,
H.~Abramowicz$^{\rm 154}$
% ...
\bigskip
\\
% ...
$^{11}$ Institute of Physics, Azerbaijan Academy of Sciences, Baku, Azerbaijan\\
% ...
$^{84}$ CPPM, Aix-Marseille Universit{\'e} and CNRS/IN2P3, Marseille, France\\
% ...
$^{89}$ Department of Physics and Astronomy, Michigan State University, East Lansing MI, United States of America\\
% ...
$^{106}$ Nikhef National Institute for Subatomic Physics and University of Amsterdam, Amsterdam, Netherlands\\
% ...
$^{112}$ Homer L. Dodge Department of Physics and Astronomy, University of Oklahoma, Norman OK, United States of America\\
$^{113}$ Department of Physics, Oklahoma State University, Stillwater OK, United States of America\\
% ...
$^{116}$ LAL, Universit{\'e} Paris-Sud and CNRS/IN2P3, Orsay, France\\
% ...
$^{152}$ Institute of Physics, Academia Sinica, Taipei, Taiwan\\
% ...
$^{154}$ Raymond and Beverly Sackler School of Physics and Astronomy, Tel Aviv University, Tel Aviv, Israel\\
% ...
$^{159}$ Department of Physics, University of Toronto, Toronto ON, Canada\\
% ...
\end{flushleft}

%\end{document}
% Created with ./xml2latex.py

\end{document}